\documentclass[11pt]{article}
\usepackage{graphicx, amssymb}
\newcommand{\SetFigFont}[3]{}

\title{A Rigorous Treatment of Energy Extraction from
a Rotating Black Hole}
\author{F.\ Finster\thanks{Research supported in part by the Deutsche
Forschungsgemeinschaft.}, N.\ Kamran\thanks{Research supported by NSERC grant
\# RGPIN 105490-2004.},
J.\ Smoller\thanks{Research supported in
part by the Humboldt Foundation and the National Science Foundation,
Grant No.~DMS-0603754.}, and S.-T.\ Yau\thanks{Research supported in part
by the NSF, Grant No.\ 33-585-7510-2-30.}}
\date{\today}

\newtheorem{Def}{Def.}[section]
\newtheorem{Thm}[Def]{Theorem}
\newtheorem{Prp}[Def]{Proposition}
\newtheorem{Lemma}[Def]{Lemma}
\newtheorem{Remark}[Def]{Remark}

\newcommand{\Proof}{{\em{Proof. }}}
\newcommand{\QED}{\ \hfill $\FBox$ \\[1em]}
\newcommand{\spc}{\;\;\;\;\;\;\;\;\;\;}
\newcommand{\bra}{\mbox{$< \!\!$ \nolinebreak}}
\newcommand{\ket}{\mbox{\nolinebreak $>$}}
\newcommand{\lbra}{\langle}
\newcommand{\lket}{\rangle}
\newcommand{\C}{\mathbb{C}}
\newcommand{\R}{\mathbb{R}}

\newcommand{\Z}{\mathbb{Z}}

\newcommand{\N}{\mathbb{N}}
\newcommand{\sN}{\mbox{\rm \scriptsize I \hspace{-.8 em} N}}

\newcommand{\beq}{\begin{equation}}
\newcommand{\eeq}{\end{equation}}

\newcommand{\FBox}{\rule{2mm}{2.25mm}}

\newcommand{\Etot}{E_{\mbox{\scriptsize{\rm{tot}}}}}
\newcommand{\Ebh}{E_{\mbox{\scriptsize{\rm{bh}}}}}
\newcommand{\Mbh}{A_{\mbox{\scriptsize{\rm{bh}}}}}
\newcommand{\Eout}{E_{\mbox{\scriptsize{\rm{out}}}}}
\newcommand{\cin}{c_{\mbox{\scriptsize{\rm{in}}}}}
\newcommand{\cout}{c_{\mbox{\scriptsize{\rm{out}}}}}

\begin{document}
\maketitle

\begin{abstract}
The Cauchy problem is considered for the scalar wave equation in the Kerr geometry.
We prove that by choosing a suitable wave packet as initial data,
one can extract energy from the black hole, thereby putting supperradiance,
the wave analogue of the Penrose process, into a rigorous mathematical framework.
We quantify the maximal energy gain. We also compute the infinitesimal change of
mass and angular momentum of the black hole, in agreement with Christodoulou's
result for the Penrose process. The main mathematical tool is our previously
derived integral representation of the wave propagator.
\end{abstract}

\section{Introduction and Statement of Results}
\setcounter{equation}{0}
Near a rotating black hole there is the counter-intuitive effect that the
physical energy of particles or waves need not be positive.
As discovered by Roger Penrose~\cite{Penrose}, this effect can be used
to extract energy from the black hole. In the process proposed by Penrose,
a classical particle enters the so-called {\em{ergosphere}}, a region outside
the event horizon, where it disintegrates into two particles. One particle
of negative energy falls into the black hole, whereas the other particle has
energy greater than the initial energy, and escapes to infinity.
The wave analogue of the Penrose process is called {\em{superradiance}};
it was proposed in~\cite{Z} and first studied in~\cite{S, TeP} and~\cite{DR1, DR2}, see
also~\cite{Ch, FN, Wald}.
In these papers, superradiance was considered only on the level of modes,
i.e., by considering the reflection coefficients for the radial ODE
obtained after separating out the time and angular dependence.
A more convincing treatment of superradiance is to consider
the Cauchy problem for ``wave packet'' initial data, and in~\cite{ALP},
this was carried out numerically for the scalar wave equation.
In this paper we shall consider the Cauchy problem analytically,
giving a mathematically rigorous treatment of superradiance for scalar waves.
Our analysis is based on the integral representation for the wave
propagator obtained in~\cite{FKSY1, FKSY2}.

A rotating black hole is modeled by the Kerr metric, which in
Boyer-Lindquist coordinates~$(t, r, \vartheta, \varphi)$ with
$r>0$, $0 \leq \vartheta \leq \pi$, $0 \leq \varphi < 2\pi$, takes the form
\begin{eqnarray}
\lefteqn{ ds^2 \;=\; g_{jk}\:dx^j\, dx^k } \nonumber \\
&=& \frac{\Delta}{U} \:(dt \:-\: a \:\sin^2 \vartheta \:d\varphi)^2
\:-\: U \left( \frac{dr^2}{\Delta} + d\vartheta^2 \right) \:-\:
\frac{\sin^2 \vartheta}{U} \:(a \:dt \:-\: (r^2+a^2) \:d\varphi)^2 .\quad
\label{eq:0}
\end{eqnarray}
Here $M>0$ and $aM>0$ denote the mass and the angular momentum of the black hole,
respectively, and
the functions~$U$ and $\Delta$ are given by
\[ U(r, \vartheta) \;=\; r^2 + a^2 \:\cos^2 \vartheta \;,\spc
\Delta(r) \;=\; r^2 - 2 M r + a^2  \: . \]
We consider only the {\em{non-extreme case}} $M^2 > a^2$, where the function $\Delta$
has two distinct roots. The largest root
\[ r_1 \;=\; M \:+\: \sqrt{M^2 - a^2} \]
defines the the event horizon of the black hole.
We restrict attention to the region $r>r_1$ outside the event horizon
where~$\Delta>0$.
The metric~(\ref{eq:0}) does not depend on~$t$ nor~$\varphi$
and is thus stationary and axisymmetric, admitting the two commuting Killing fields
$\frac{\partial}{\partial t}$ and $\frac{\partial}{\partial \varphi}$.
The {\em{ergosphere}} is defined to be the region where the Killing
field $\frac{\partial}{\partial t}$ is space-like, that is where
\begin{equation}
r^{2}-2Mr+a^{2}\,\cos^{2}\vartheta < 0\:.
\end{equation}
The ergosphere lies outside the event
horizon $r=r_{1}$, and its boundary intersects the event
horizon at the poles $\vartheta=0,\,\pi$.

We now briefly review the Penrose process (for more details see~\cite{Wald}).
The $4$-momentum $p^j$ of a massive point particle is time-like and
future-directed, and thus its energy $\langle p, \frac{\partial}{\partial t} \rangle$ is
positive whenever~$\frac{\partial}{\partial t}$ is time-like.
However, in the ergosphere the vector~$\frac{\partial}{\partial t}$ becomes space-like,
and hence the energy of the point particle need {\em{not}} be positive.
Penrose considers a particle of positive energy which splits inside the
ergosphere into two particles whose energies have opposite signs.
By finely tuning the energy and momenta of these particles, one can arrange
that the particle of negative energy crosses the event horizon and reduces
the angular momentum of the black hole, whereas the other particle escapes to
infinity, carrying (due to energy conservation) more energy than the original
particle. In this way, one can extract energy from the black hole, at the cost of
reducing its angular momentum. Christodoulou~\cite{Christo} showed that
the infinitesimal changes of mass $\delta M$ and angular momentum $\delta(aM)$ of the black hole
satisfy the inequalities
\beq \label{deltaM}
\delta (aM) \;\leq\; \frac{r_1^2+a^2}{a} \:\delta M \;<\; 0\:,
\eeq
and as a consequence he showed that it is not possible to reduce the mass of the
black hole via the Penrose process below the {\em{irreducible mass}}
$M_{\mbox{\scriptsize{irr}}}$ defined by
\beq \label{Mirr}
M_{\mbox{\scriptsize{irr}}}^2 \;=\; \frac{1}{2} \left( M^2 + \sqrt{M^4 -(aM)^2} \right)\:.
\eeq

We now recall the wave equation in the Kerr geometry and its separation;
for more details see~\cite{FKSY1}. The scalar wave equation is
\begin{equation} \label{swave}
g^{ij}\nabla_{i}\nabla_{j}\,\Phi=\frac{1}{\sqrt {-\det g}}\,\frac{\partial}{\partial
x^{i}} \left( {\sqrt{-\det g}}\,g^{ij}\frac{\partial}{\partial
x^{j}} \right) \Phi \;=\; 0 \:,
\end{equation}
and in Boyer-Lindquist coordinates this becomes
\begin{eqnarray} \label{wave}
\lefteqn{ \left[ -\frac{\partial}{\partial r}\Delta\frac{\partial}{\partial r}
+\frac{1}{\Delta} \left( (r^{2}+a^{2})\frac{\partial}{\partial
t}+a\frac{\partial}{\partial \varphi} \right)^{2}
-\frac{\partial}{\partial \cos \vartheta} \sin^{2}\vartheta
\frac{\partial}{\partial \cos \vartheta} \right. \nonumber } \\
&& \hspace*{5.5cm} \left. -\frac{1}{\sin^{2}\vartheta} \left( a\sin^{2}\vartheta
\frac{\partial}{\partial t}+\frac{\partial}{\partial
\varphi}\right)^{2} \right ]\Phi \;=\; 0\:.
\end{eqnarray}
Using the ansatz
\begin{equation}\label{separansatz}
\Phi(t,r,\vartheta,\varphi)=e^{-i\omega
t-ik\varphi}\:R(r)\:\Theta(\vartheta)
\end{equation}
with~$\omega \in \R$ and~$k \in \Z$,
the wave equation can be separated into both angular and radial ODEs,
\begin{equation} \label{odes}
{\cal{R}}_{\omega ,k}\,R_\lambda \;=\; -\lambda \, R_{\lambda},\qquad  {\cal{A}}_{\omega ,k}\,\Theta_\lambda \;=\; \lambda \, \Theta_\lambda\:.
\end{equation}
The angular operator~${\cal{A}}_{\omega, k}$ is also called the
{\em{spheroidal wave operator}}. It has a purely discrete spectrum
of non-degenerate eigenvalues~$0 \leq \lambda_1 < \lambda_2,
\ldots$. The corresponding eigenfunctions~$\Theta^{\omega,k}_n$ are
referred to as spheroidal wave functions. In order to bring the
radial equation into a convenient form, we introduce a new radial
function $\phi(r)$ by
\begin{equation}\label{rescal}
\phi(r) \;=\; \sqrt{r^2+a^2}\: R(r)\:, \end{equation}
and define the
Regge-Wheeler variable~$u \in \R$ by
\begin{equation} \label{51a}
\frac{du}{dr} \;=\; \frac{r^2+a^2}{\Delta} \:,
\end{equation}
mapping the event horizon to $u=-\infty$. The radial equation then
takes the form of the Schr{\"o}dinger equation
\begin{equation}
\left(-\frac{d^2}{d u^2} + V(u) \right) \phi(u) \;=\; 0
\label{5ode}
\end{equation}
with the potential
\begin{equation} \label{5V}
V(u) \;=\; -\left( \omega + \frac{ak}{r^2+a^2} \right)^2 \:+\:
\frac{\lambda_n(\omega)\:\Delta}{(r^2+a^2)^2} \:+\: \frac{1}{\sqrt{r^2+a^2}}\; \partial_u^2 \sqrt{r^2+a^2} \:.
\end{equation}

Starobinsky~\cite{S} analyzes superradiance on the level of modes. Using the
notation in~\cite{FKSY2}, we can explain his ideas and results as follows.
We fix the separation constants~$k>0$, $\omega$ and $\lambda_n$.
Introducing the notation
\beq \label{Odef}
\Omega \;=\; \omega - \omega_0 \spc {\mbox{with}} \spc
\omega_0 \;=\; -\frac{ak}{r_1^2+a^2}\:,
\eeq
the potential~(\ref{5V}) has the asymptotics
\[ \lim_{u \to -\infty} V(u) \;=\; -\Omega^2 \:,\spc
\lim_{u \to \infty} V(u) \;=\; -\omega^2\:. \]
Thus there are fundamental solutions~$\acute{\phi}$ and~$\grave{\phi}$
of~(\ref{5ode}) which behave asymptotically like plane waves,
\begin{eqnarray}
\lim_{u \to -\infty} e^{-i \Omega u} \:\acute{\phi}(u) &=& 1 \:,\spc
\lim_{u \to -\infty} \left(e^{-i \Omega u} \:\acute{\phi}(u) \right)' \;=\; 0  \label{abc1} \\
\lim_{u \to \infty} e^{i \omega u} \:\grave{\phi}(u) &=& 1 \:,\spc\;\;\;\;\,
\lim_{u \to \infty} \left(e^{i \omega u} \:\grave{\phi}(u) \right)' \;=\; 0\:,  \label{abc2}
\end{eqnarray}
(see~\cite{FKSY2} for details). The solution~$\phi =
\overline{\acute{\phi}}$ has, near the event horizon $u=-\infty$,
the asymptotics $\phi(u) = e^{-i \Omega u}$. Taking into account the
factor~$e^{-i \omega t}$ in the separation, this corresponds to a
plane wave entering the black hole. A general solution~$\phi$ can be
expressed as a linear combination of~$\grave{\phi}$ and its complex
conjugate,
\begin{equation}\label{decompphi}
\phi(u) \;=\; A\:
\grave{\phi}(u) + B \overline{\grave{\phi}}(u) \:.
\end{equation}
Equation~(\ref{abc2}) shows that the
solution~$\grave{\phi}$ behaves near infinity like $\grave{\phi}(u)
= e^{-i \omega u}$. Hence the corresponding time-dependent solution
behaves like the plane wave~$e^{-i \omega(t+u)}$ and
corresponds to an {\em{incoming wave}} propagating from infinity.
Likewise, the
solution~$\overline{\grave{\phi}}$ corresponds to an {\em{outgoing
wave}} propagating towards infinity. Computing the Wronskian
of~$\phi$ and $\overline{\phi}$ near the event horizon and near
infinity gives the relation
\beq \label{ABeq} |A|^2 - |B|^2 \;=\; \frac{\Omega}{\omega} \:.
\eeq
The quantities~$\omega^2 |A|^2$
and~$\omega^2 |B|^2$ have the interpretations as the energy flux of
the incoming and outgoing waves, respectively. If the right side
of~(\ref{ABeq}) is positive, the outgoing flux is smaller than the
incoming flux, and this corresponds to ordinary scattering. However,
if the right side of~(\ref{ABeq}) is negative, then the outgoing
flux is larger than the incoming flux; this is termed
{\em{superradiant scattering}}. According to~(\ref{ABeq}),
superradiant scattering appears precisely when $\omega$ and $\Omega$
have opposite signs. Using~(\ref{Odef}), we see that superradiant
scattering occurs if and only if $\omega$ lies in one of the following intervals
\beq
\label{omegasuper} \omega_0 \;<\; \omega \;<\; 0 \:,\spc
0 \:<\; \omega \;<\; \omega_0\:,
\eeq
depending on the sign of~$\omega_0$. The gain in energy is determined
by the quotient of outgoing and incoming flux,
\beq \label{frakRdef}
{\mathfrak{R}} \;=\; \frac{|B|^2}{|A|^2} \:.
\eeq
Starobinsky~\cite{S} made an asymptotic expansion
for~${\mathfrak{R}}$ and found a relative gain of energy of
about~$5\%$ for $k=1$ and less than~$1\%$ for $k \geq2$. This
was verified later numerically in~\cite{ALP}. Teukolsky and
Press~\cite{TeP} made a similar mode analysis for higher spin and
found numerically an energy gain of at most~$4.4\%$ for Maxwell
($s=1$) and up to $138\%$ for gravitational waves ($s=2$).

Our main result makes the above mode argument for the scalar wave equation
mathematically precise in the setting of the Cauchy problem.
To state our result, we combine~$\Phi$ and $\partial_t\Phi$,
as in~\cite{FKSY1}, into a two-component vector~$\Psi = (\Phi, i\partial_t \Phi)$.
We always assume that the initial data is smooth and
compactly supported outside the event horizon,
\[ \Psi_0 \;=\; (\Phi, i \partial_t \Phi)|_{t=0} \;\in\; C^\infty_0((r_1, \infty) \times S^2)^2\:. \]
The physical energy of the wave is then given by~$\bra \Psi, \Psi \ket$, where
$\bra .,. \ket$ is the {\em{energy scalar product}}
\begin{eqnarray}\label{energyprod}
\lefteqn{ \bra \Psi, \Psi'\ket \;=\;
\int_{r_1}^{\infty}dr\int_{-1}^{1} d(\cos \vartheta)  \left\{
\left({\frac{(r^{2}+a^{2})^{2}}{\Delta}} - a^{2}\,\sin^{2}\vartheta \right) \overline{\partial_{t} \Phi}\, \partial_{t}{\Phi}' \right. \nonumber } \\
&& \left. +\Delta \, \overline{\partial_{r}\Phi}
\,\partial_{r}{\Phi}' +\sin^{2}\vartheta
\,\overline{\partial_{\cos \vartheta}\Phi}\,\partial_{\cos
\vartheta}{\Phi}'  +\left(
{\frac{1}{\sin^{2}\vartheta}}-{\frac{a^{2}}{\Delta}}\right)\,\overline{\partial_{\varphi}\Phi}
\,\partial_{\varphi}{\Phi}' \right\}. \label{energysp}
\end{eqnarray}
Provided that the limit $t\rightarrow \infty$ exists,
the energy radiated to infinity can be defined by
\beq \label{Eoutdef}
\Eout \;=\; \lim_{t \to \infty} \bra \Psi(t), \chi_{(2r_1, \infty)}(r)\, \Psi(t)\ket \:,
\eeq
where $\chi$ is the characteristic function. Note that,
according to the pointwise decay result in~\cite{FKSY2}, we could replace
the argument~$2r_1$ by any radius greater than~$r_1$. Moreover, again using pointwise local decay,
the boundary term at $r=2 r_1$ which arises when differentiating the characteristic
function~$\chi_{(2r_1, \infty)}$, vanishes in the limit~$t \rightarrow \infty$.
We can now state our main theorem.

\begin{Thm}\label{maintheo} Consider the Cauchy problem for the scalar wave equation in
the non-extreme Kerr geometry for initial data with compact support
outside the event horizon,
\[ \Psi_0 \;=\; (\Phi, i \partial_t \Phi)|_{t=0} \;\in\; C^\infty_0((r_1, \infty) \times S^2)^2\:, \]
having energy~$\bra \Psi_0, \Psi_0 \ket$.
Fix $k>0$, $n \in \N$ and~$\omega \in (\omega_0,0)$.
Then for any~$R>r_1$
and~$\delta>0$ there is initial data $\Psi_0 \in C^\infty_0((R, \infty) \times S^2)^2$ such that
the limit in~(\ref{Eoutdef}) exists and
\[ \left| \frac{\Eout}{\bra \Psi_0, \Psi_0 \ket}
- {\mathfrak{R}} \right| \;\leq\; \delta \]
with~$\mathfrak{R}$ as in~(\ref{frakRdef}). The same inequality holds
for~$k<0$ and~$\omega \in (0, \omega_0)$.
\end{Thm}
We emphasize that we allow the initial data to be supported arbitrarily far away
from the event horizon. This is important in order to avoid artificial initial
data which would not correspond to an energy extraction mechanism. For example, if one
allows the support of the initial data to intersect the ergosphere, one could
take initial data close to a wave packet with zero total energy, in which case the quotient~$\Eout/\bra
\Psi_0, \Psi_0 \ket$ could be made arbitrarily large.

We also compute the energy~$\Ebh$ and the angular momentum~$\Mbh$ of the
wave component crossing the event horizon of the black hole to obtain
the following theorem.
\begin{Thm} \label{thm2}
For any initial data~$\Psi_0 \in C^\infty_0((r_1, \infty) \times S^2)^2$,
the quantities~$\Ebh$ and~$\Mbh$ defined by~(\ref{Eindef}, \ref{Mbhdef}) satisfy
the inequality
\[ \Mbh \;\leq\; \frac{r_1^2+a^2}{a} \: \Ebh\:. \]
\end{Thm}
This shows explicitly that Christodoulou's estimates~(\ref{deltaM}, \ref{Mirr})
also hold for energy extraction by scalar waves,
in agreement with Hawking's area theorem~\cite{Wald}.

\section{Absorbtion of Energy by the Black Hole}
\setcounter{equation}{0}
Recall from \cite{FKSY2} that given initial data $\Psi_0 \in
C^\infty_0(\R \times S^2)^2$, the solution of the Cauchy problem
for the scalar wave equation (\ref{wave}) can be represented as
\begin{equation}\label{intrep} \Psi(t,r,\vartheta,
\varphi) \;=\; \frac{1}{2 \pi} \sum_{k \in \Z} e^{-i k \varphi}
\sum_{n \in \sN} \int_{-\infty}^\infty \frac{d\omega}{\omega
\Omega}\:e^{-i \omega t} \sum_{a,b=1}^2 t^{k \omega n}_{ab}\:
\Psi^a_{k \omega n}(r,\vartheta)\; \bra \Psi^b_{k \omega n}, \Psi_0
\ket\:,
\end{equation}
where the sums and the integrals converge
in~$L^2_{\mbox{\scriptsize{loc}}}$. In the above integral
representation, the coefficients $t_{ab}$ are defined by \beq
\label{tabdef} t_{11} \;=\; 1 +
{\mbox{\rm{Re}}}\,\frac{\alpha}{\beta} \:,\qquad t_{12} \;=\;
t_{21} \;=\; -{\mbox{\rm{Im}}}\, \frac{\alpha}{\beta} \:,\qquad
t_{22} \;=\; 1 - {\mbox{\rm{Re}}}\, \frac{\alpha}{\beta}\:, \eeq
and the complex coefficients $\alpha$ and $\beta$ are defined by
\beq \label{trans} \grave{\phi}
\;=\; \alpha\: \acute{\phi} + \beta\: \overline{\acute{\phi}} \:,
\eeq
where~${\acute{\phi}}(u)$ and~${\grave{\phi}}(u)$ are the fundamental solutions
of the radial equation having the asymptotics~(\ref{abc1}) and~(\ref{abc2}), respectively.
 We will refer to the complex coefficients $\alpha$ and
$\beta$ as {\em{transmission coefficients}}. Finally the functions
$\Psi^a_{k \omega n}(r,\vartheta),\,a=1,2$ are the solutions of
the wave equation (\ref{wave}), with fixed quantum numbers
$k,\omega,n $, corresponding to the real-valued fundamental
solutions of the radial equation given by
\beq \label{phi12}
\phi^{\,1}={\mbox{Re}}\,{\acute{\phi}}\;,\qquad
\phi^{\,2}={\mbox{Im}}\, {\acute{\phi}}\:.
\eeq
Here we shall always consider a fixed $k$-mode, and without loss of
generality we assume that~$k>0$. For notational simplicity, from now on
we omit the index~$k$ and the~$\varphi$-dependence.
Furthermore, we introduce the vector notation
\beq \label{vn}
{\bf{\Psi}}^{\omega n}(r,\vartheta) \;:=\;
\left(\!\begin{array}{cc} \Psi_{1}^{\omega n}(r,\vartheta) \\
\Psi_{2}^{\omega n}(r,\vartheta) \end{array}\!\right) , \quad
{\bf{\Psi}}_{0}^{\omega n} \;:=\;
\left(\!\begin{array}{cc} \bra \Psi_{1}^{\omega n},\Psi_{0}\ket \\
\bra \Psi_{2}^{\omega n},\Psi_{0}\ket
\end{array}\!\right) , \qquad
{\bf{T}} \;=\; (t_{ab})_{a,b=1,2} \:,
\eeq
which allows us to write the integral representation~(\ref{intrep})
in the compact form
\beq \label{intrep2}
\Psi(t,r,\vartheta) \;=\; \frac{1}{2 \pi}
\sum_{n \in \sN} \int_{-\infty}^\infty \frac{d\omega}{\omega
\Omega}\:e^{-i \omega t}
\langle {\bf{\Psi}}^{\omega n}(r,\vartheta),\, {\bf{T}}\,
{\bf{\Psi}}^{\omega n}_0 \rangle_{\C^2} \:.
\eeq

We now introduce some basic quantities needed for the
formulation of the superradiance property for wave packets. The {\em
total energy} $\Etot$ of an initial data
~$\Psi_0 \in C^\infty_0(\R \times S^2)^2$ for the scalar wave
equation (\ref{wave}) is defined by
\begin{equation}\label{Etotdef}
\Etot \;=\;\bra \Psi_0,\Psi_0 \ket\;.
\end{equation}
This energy is conserved throughout the time evolution of the
scalar wave, meaning that if $\Psi(t)$ denotes the solution of the
Cauchy problem for the scalar wave equation, then
\[
\bra \Psi_0,\Psi_0 \ket\;=\;\bra \Psi(t),\Psi(t) \ket\;.
\]
Recall that the {\em outgoing energy}, which represents the energy radiated to
infinity, has been defined in~(\ref{Eoutdef}).
Finally, the energy absorbed by the black hole is defined by
\beq \label{Eindef}
\Ebh \;=\; \lim_{t \to
\infty} \bra \Psi(t), \chi_{(-\infty, 2r_{1})}(r)\, \Psi(t)\ket
\:. \eeq

We next derive useful expressions for these quantities in
terms of the initial data and the transmission coefficients
$\alpha$ and $\beta$. The expression for~$\Etot$ is an immediate
consequence of~(\ref{intrep2}).

\begin{Prp}\label{Etotexpr}
Choose a fixed $k\in \Z$ such that $\omega_{0}<0$. Then
\[ \Etot \;=\;
\frac{1}{2 \pi}\sum_{n \in \sN}\int_{-\infty}^{\infty}\frac{d\omega}{\omega \Omega}\:
\langle {\bf{\Psi}}^{\omega n}_0, {\bf{T}}\, {\bf{\Psi}}^{\omega n}_0
\rangle_{\C^2} , \]
where the series converges absolutely.
\end{Prp}

We next want to compute~$\Ebh$ and~$\Eout$.
First, we need an argument which ensures that the sum over the angular momentum
modes converges. Since $\Eout = \Etot - \Ebh$ and the convergence
of the angular sum is already known for the total energy
according to Proposition~\ref{Etotexpr}, it suffices to consider
for example~$\Eout$. In~\cite{FS} it is shown that the
outgoing energy is bounded uniformly in time, i.e.
\[ \bra \Psi(t), \chi_{(2r_1, \infty)}(r)\, \Psi(t)\ket \;\leq\;
C(\Psi) \qquad {\mbox{for all~$t$}}\:. \]
Moreover, it is also shown in~\cite{FS} that the outgoing wave can be
approximated by a finite number of angular momentum modes
uniformly in time, i.e.\ for any~$\delta>0$ there is~$N$ such that
\[ \bra \Psi^N(t), \chi_{(2r_1, \infty)}(r)\, \Psi^N(t)\ket \;\leq\;
\delta \qquad {\mbox{for all~$t$}} \:, \]
where
\[ \Psi^N(t,r,\vartheta,
\varphi) \;=\; \frac{1}{2 \pi}
\sum_{n \geq N} \int_{-\infty}^\infty \frac{d\omega}{\omega
\Omega}\:e^{-i \omega t} \sum_{a,b=1}^2 t^{\omega n}_{ab}\:
\Psi_a^{\omega n}(r,\vartheta)\; \bra \Psi_b^{\omega n}, \Psi_0
\ket\:. \]
This estimate will always allow us in what follows to
interchange the limit~$t \rightarrow \infty$ with the
sum over the angular momentum modes.

To compute~$\Ebh$, we make use of the following lemma, which
generalizes~\cite[Lemma~9.1]{FKSY02}.
\begin{Lemma}\label{Fourierprod}
For any Schwartz function $f \in {\mathcal S}(\R \times \R)$ and any~$u_0 \in \R$,
we define the four functions~$A_{s,s'}$, where~$s,s' \in \{-1,1\}$, by
\[ A_{s,s'} \;=\; \lim_{t\to \infty} \int_{-\infty}^{u_{0}}du
\int_{-\infty}^{\infty}d\omega
\int_{-\infty}^{\infty}d\omega' e^{-i(\omega-\omega')t}
\: e^{-i s \Omega u + i s' \Omega' u}
f(\omega,\omega') \:. \]
Then
\[ A_{s,s'} \;=\; 2\pi \left( \delta^s_1 \: \delta^{s'}_1 \right)
\int_{-\infty}^{\infty} f(\omega,\omega)\: d\omega\:, \]
where~$\delta^s_1$ denotes the Kronecker delta.
\end{Lemma}
{\Proof} Noting that~$\Omega-\Omega'=\omega-\omega'$, the case~$s=s'$ was
proved in~\cite[Lemma~9.1]{FKSY02}. Thus it remains to show that
\[ \lim_{t\to \infty} \int_{-\infty}^{u_{0}}du
\int_{-\infty}^{\infty}d\omega
\int_{-\infty}^{\infty}d\omega' e^{-i(\omega-\omega')t}
\: e^{\pm i (\Omega + \Omega') u}
f(\omega,\omega') \;=\; 0 \:.\]
Noting
\[ e^{-i(\omega-\omega')t}
\: e^{\pm i (\Omega + \Omega') u} \;=\; \frac{1}{(-it \pm iu+1)(it\pm iu+1)}\:
(\partial_\omega+1) (\partial_{\omega'}+1)
e^{-i(\omega-\omega')t \pm i (\Omega + \Omega') u} \:, \]
we can integrate the~$\omega,\omega'$-derivatives by parts. Applying Fubini's theorem
and estimating the resulting $u$-integral for~$t>2|u_0|$, we find
\begin{eqnarray*}
\left| \int_{-\infty}^{u_0} \frac{1}{(t \mp u+i)(t\pm u-i)}\:
e^{\pm i (\Omega + \Omega') u}\: du \right| &=&
\int_{-\infty}^{u_0} \left((t+u)^2+1 \right)^{-\frac{1}{2}}
 \left((t-u)^2+1 \right)^{-\frac{1}{2}}\: du \\
 &\leq& \sqrt{6}\:t^{-\frac{1}{2}} \int_{-\infty}^{u_0}
 \left((t+u)^2+1 \right)^{-\frac{3}{4}}\: du\:.
\end{eqnarray*}
Since the last integral is bounded uniformly in~$t$, the entire expression tends to zero
as $t \rightarrow \infty$.
\QED

\begin{Prp}\label{Einexpr}
Choose a fixed integer~$k>0$. Then the limit in~(\ref{Eindef}) converges and
\[ \Ebh \;=\; \frac{1}{2 \pi}\sum_{n \in \sN}\int_{-\infty}^{\infty}\frac{d\omega}{\omega \Omega}\:
\lbra {\bf{\Psi}}_{0}^{\omega n},{\bf T}{\bf P}{\bf T} \,{\bf{\Psi}}_{0}^{ \omega
n}\lket_{\C^{2}}\:,\]
where~${\bf{P}}$ is the matrix
\begin{eqnarray*}
{\bf P}\;=\; \frac{1}{2} \left(\!\begin{array}{cc}1 & i \\
-i & 1
\end{array}\! \right) .
\end{eqnarray*}
The above series is absolutely convergent.
\end{Prp}
{\Proof} Substituting the integral representation~(\ref{intrep}) into~(\ref{Eindef}),
we find
\beq \label{Ebhex}
\Ebh \;=\; \lim_{t\to
\infty}\frac{1}{(2\pi)^{2}}
\int_{r_1}^{2 r_1} dr \int_{-\infty}^{\infty}\frac{d\omega'}{\omega' \Omega'}
\int_{-\infty}^{\infty}\frac{d\omega}{\omega \Omega} \:e^{-i (\omega-\omega') t}
\:{\mathcal{E}}^{\omega', \omega}(r) \;,
\eeq
where
\[ {\mathcal{E}}^{\omega', \omega}(r) \;=\; \sum_{n,n' \in \N}
\sum_{a,b,c,d=1}^2 t^{\omega' n'}_{ab} t^{\omega n}_{cd}\:
\bra \Psi_0, \Psi^{\omega' n'}_a \ket
\: \bra \Psi^{\omega n}_d, \Psi_0 \ket
\int_{S^2} {\mathcal{E}}(\Psi^{\omega' n'}_b, \Psi^{\omega n}_c)\:, \]
and~${\mathcal{E}}(\Psi^{\omega' n'}_b, \Psi^{\omega n}_{c})$ denotes the
bilinear form corresponding to the energy density as given in~(\ref{energysp}).

We next justify that we may interchange the limit~$t \rightarrow \infty$ with the
sums over the angular momentum modes and that the series converge absolutely.
According to~\cite[Theorem~8.1]{FS},
the energy of the large angular momentum modes near infinity is small
uniformly in time. Due to conservation of energy and the local decay~\cite{FKSY2}, 
a similar result follows for the energy near the event horizon.
Hence the integrals in~(\ref{Ebhex}) can be approximated uniformly in time
by a finite number of angular momentum modes. Thus in what follows we may
restrict attention to fixed angular momentum modes~$n$ and~$n'$.

Near the event horizon, we can expand the coefficient functions in the
energy density~${\mathcal{E}}(\Psi^{\omega' n'}_b, \Psi^{\omega n}_{c})$ to obtain
\begin{eqnarray*}
\lefteqn{ \frac{\Delta}{r^2+a^2}\: {\mathcal{E}}(\Psi^{\omega' n'}_b, \Psi^{\omega n}_{c}) } \\
&=& (r_1^2+a^2)
\left( \omega' \omega\:\overline{\Phi^{\omega' n'}_b}\, \Phi^{\omega n}_c
+ \overline{\partial_u \Phi^{\omega' n'}_b}\,\partial_u \Phi^{\omega n}_c
\right) \:-\: \frac{a^2 k^2}{r_1^2+a^2}\:\overline{\Phi^{\omega' n'}_b}\, \Phi^{\omega n}_c
\:+\: {\mathcal{O}}_{\omega,\omega'}(e^{\gamma u}) \\
&=& \Theta_{n'\omega'} \Theta_{n \omega}
\left( \omega' \omega\:\overline{\phi^{\omega' n'}_b}\, \phi^{\omega n}_c
+ \overline{\partial_u \phi^{\omega' n'}_b}\,\partial_u \phi^{\omega n}_c
\:-\: \frac{a^2 k^2}{(r_1^2+a^2)^2}\:\overline{\phi^{\omega' n'}_b}\, \phi^{\omega n}_c \right)
\:+\: {\mathcal{O}}_{\omega,\omega'}(e^{\gamma u})
\end{eqnarray*}
with the fixed constant~$\gamma>0$ as in~\cite[Eq.~(3.9)]{FKSY2}. Here the error term
can be bounded by
\beq \label{erroro}
|{\mathcal{O}}_{\omega,\omega'}(e^{\gamma u}) | \;\leq\; C\, e^{\gamma u}
\left( |\overline{\phi^{\omega' n'}_b}| + |\omega' \overline{\phi^{\omega' n'}_b}| +
|\partial_u \overline{\phi^{\omega' n'}_b}| \right)
\left( |\overline{\phi^{\omega n}_a}| + |\omega \overline{\phi^{\omega n}_a}| +
|\partial_u \overline{\phi^{\omega n}_a}| \right)
\eeq
with a constant~$C$ independent of $\omega$ and~$\omega'$.
Using the Jost estimates of~\cite[Theorem~3.1]{FKSY2}, the fundamental solutions
have the following expansion near the event horizon,
\[ \phi^{\omega n}_1(u) \;=\; \cos(\Omega u) \:+\: \tilde{\mathcal{O}}_{\omega,\omega'}(e^{\gamma u})\:,\quad \phi^{\omega n}_2(u) \;=\; \sin(\Omega u) \:+\: \tilde{\mathcal{O}}_{\omega,\omega'}(e^{\gamma u}) \:. \]
Here the error term is bounded by
\[ | \tilde{\mathcal{O}}_{\omega,\omega'}(e^{\gamma u}) | \;\leq\; C\, e^{\gamma u}\:,\qquad
| \partial_u \tilde{\mathcal{O}}_{\omega,\omega'}(e^{\gamma u}) | \;\leq\; C\,(1+|\omega|) e^{\gamma u}\:, \]
where the constant~$C$ is again uniform in~$\omega$ (see~\cite[eqns~(3.7, 3.10)]{FKSY2}).
This estimate also shows that the error term~(\ref{erroro}) is bounded by
\beq \label{erroro2}
| {\mathcal{O}}_{\omega,\omega'}(e^{\gamma u}) | \;\leq\; C\, e^{\gamma u}\: (1+|\omega'|)
(1+|\omega|)\:.
\eeq
We thus conclude that
\begin{eqnarray}
\lefteqn{ \frac{\Delta}{r^2+a^2}\: {\mathcal{E}}(\Psi^{\omega' n'}_b, \Psi^{\omega n}_{c}) 
\;=\; {\mathcal{O}}_{\omega,\omega'}(e^{\gamma u}) } 
\nonumber \\
&& +\:\Theta_{n'\omega'} \Theta_{n \omega}
\left( \omega' \omega\:\overline{\psi^{\omega'}_b}\, \psi^{\omega}_c
+ \overline{\partial_u \psi^{\omega'}_b}\,\partial_u \psi^{\omega}_c
\:-\: \frac{a^2 k^2}{(r_1^2+a^2)^2}\:\overline{\psi^{\omega'}_b}\, \psi^{\omega}_c \right) ,
\label{Eexp}
\end{eqnarray}
where~${\mathcal{O}}_{\omega,\omega'}(e^{\gamma u})$ satisfies~(\ref{erroro2}) (possibly with
a new constant~$C$) and
\[ \psi^{\omega}_1(u) \;=\; \cos(\Omega u) \:,\quad
\psi^{\omega}_2(u) \;=\; \sin(\Omega u) \:. \]
Using the asymptotic formula~(\ref{Eexp})
in~(\ref{Ebhex}) and writing the radial integral in terms of
the Regge-Wheeler variable~$u$, the factor~$\Delta/(r^2+a^2)$ drops out.
Applying~\cite[Lemma~3.1]{FS}, we can write~$\Ebh$ as
\begin{eqnarray*}
\Ebh &=& \lim_{t \rightarrow \infty} \int_{-\infty}^{u_0} du
\int_{-\infty}^\infty d\omega' \int_{-\infty}^\infty d\omega\: e^{-i (\omega-\omega') t} \\
&& \times 
\left( \sum_{b,c=1}^2 f_{b,c}(\omega, \omega')\: \overline{\psi^{\omega'}_b}(u)\,
\psi^{\omega}_c(u) + g(\omega,\omega') \, {\mathcal{O}}_{\omega,\omega'}(e^{\gamma u})
\right) ,
\end{eqnarray*}
with Schwartz functions $f_{b,c}$ and~$g$. To treat the error term, we first apply
Fubini's theorem to exchange the orders of integration. This gives
\[ \lim_{t \rightarrow \infty} \int_{-\infty}^\infty d\omega' \int_{-\infty}^\infty d\omega\: e^{-i (\omega-\omega') t} \:g(\omega,\omega')\: \int_{-\infty}^{u_0}
{\mathcal{O}}_{\omega,\omega'}(e^{\gamma u}) \: du\: , \]
and using~(\ref{erroro2}), we can apply the Riemann-Lebesgue to conclude that this
limit is zero. 
The term involving the~$f_{b,c}$ can be treated by applying Lemma~\ref{Fourierprod}.
Using the orthogonality of the spheroidal wave functions, we obtain
\[ \Ebh \;=\; \frac{1}{2\pi}
\int_{-\infty}^{\infty}\frac{d\omega}{\omega^2 \Omega^2}
\sum_{n \in \sN}
\lbra {\bf{\Psi}}_{0}^{\omega n},
\frac{{\bf T}{\bf P}{\bf T}}{2} \,{\bf{\Psi}}_{0}^{ \omega n}
\lket_{\C^{2}} \left( \omega^2 +\Omega^2 - \frac{a^2 k^2}{(r_1^2+a^2)^2} \right) . \]
We finally apply the identity
\[ \omega^2 +\Omega^2 - \frac{a^2 k^2}{(r_1^2+a^2)^2} \;=\;
\omega^2 +\Omega^2 - \omega_0^2 \;=\; \omega^2 + \omega^2 - \omega \omega_0 \;=\; 2 \omega \Omega\:.
\]

\vspace*{-.75cm}
\QED

Using the matrix identity
\[ {\bf{T}} - {\bf{T}} {\bf{P}} {\bf{T}} \;=\; {\bf{T}} {\bf{Q}} {\bf{T}}
\quad {\mbox{with}} \quad {\bf{Q}} := {\bf{T}}^{-1} - {\bf{P}} \:, \]
a short calculation yields the following result.
\begin{Prp}\label{Eoutexpr}
Choose a fixed $k\in \Z$ such that $\omega_{0}<0$. Then the limit~$t \rightarrow \infty$
in~(\ref{Eoutdef}) exists and
\[ \Eout \;=\; \frac{1}{2 \pi}\sum_{n \in \sN}\int_{-\infty}^{\infty}\frac{d\omega}{\omega \Omega}
\lbra {\bf{\Psi}}_{0}^{\omega n},{\bf T} {\bf Q}{\bf T} \,{\bf{\Psi}}_{0}^{
\omega n}\lket_{\C^{2}}\;,\]
where~${\bf{Q}}$ is the matrix
\begin{eqnarray*}
{\bf Q}\;=\; \frac{\Omega}{2 \omega} \left(\!\begin{array}{cc} |\alpha-\beta|^2 &
i (\alpha+\beta) \,\overline{(\alpha-\beta)} \\
-i (\alpha-\beta)\, \overline{(\alpha+\beta)}
 & |\alpha+\beta|^2 \end{array}\! \right) .
\end{eqnarray*}
\end{Prp}

\section{Energy Propagation of Wave Packets near Infinity}
\setcounter{equation}{0}
We fix ${\tilde{\omega}}\in (\omega_{0},0)$ and
${\tilde n}\in \N$ and consider initial data $\Psi_{0}$ in the
form of a linear combination of outgoing and incoming wave packets,
\beq \label{wavepack}
\Psi_{0} \;=\; \left( \!\begin{array}{cc} \Phi \\
i \partial_{t}\Phi \end{array}\! \right) \!\!\Big|_{t=0}
\;=\; \Theta_{{\tilde n},{\tilde
\omega}}(\vartheta)\: \frac{\eta_{L}(u)}{\sqrt{r^{2}+a^{2}}}
\left[ \cin\,e^{-i{\tilde \omega}u}  \left( \!\begin{array}{cc}
1 \\ {\tilde \omega} \end{array}\! \right)
+ \cout\,e^{i{\tilde \omega}u}  \left( \!\begin{array}{cc}
1 \\ -{\tilde \omega} \end{array}\! \right) \right] \:,
\eeq
where $L>0$,
\[ \eta_{L}(u)\;=\;\frac{1}{\sqrt L}\,\eta \!\left(\frac{u-L^2}{L} \right) , \]
with $\eta \in C^\infty_0((-1,1))$ being a smooth cut-off function, and where
$\Theta_{{\tilde n},{\tilde \omega}}(\vartheta)$ is an
eigenfunction of the angular operator ${\mathcal A}$.

In the next proposition we compute~$\Etot$ asymptotically as~$L \rightarrow \infty$.
\begin{Prp} \label{prpEtot}
For the initial data given by the wave packet~(\ref{wavepack}),
\beq \label{Etotprp} \lim_{L \rightarrow \infty} \Etot(\Psi_0)
\;=\; \frac{1}{4 \pi} \sum_{n \in \sN} \lim_{L \rightarrow \infty}
\int_{-\infty}^{\infty}\frac{d\omega}{\omega^2}\: \langle
\grave{\bf{\Psi}}^{\omega n}_0, \grave{\bf{\Psi}}^{\omega n}_0
\rangle_{\C^2}\:,
\eeq
where
\beq \label{psigravcomp}
\grave{{\bf{\Psi}}}^{\omega n}_0 \;:=\; 
\left(\!\begin{array}{cc} \bra \grave{\Psi}_{\omega n},\Psi_{0}\ket \\
\bra \overline{\grave{\Psi}_{\omega n}},\Psi_{0}\ket \end{array}\!
\right) . \eeq
The above series converges absolutely.
\end{Prp}
{\Proof} Applying~\cite[Lemma~8.3]{FS} with~$f \equiv 1$, we see that
the energy of the large angular momentum modes is small uniformly in~$L$.
This allows us to interchange the limit~$L \rightarrow \infty$ with the
sum over the angular momentum modes. Hence in what follows
we may restrict attention to a fixed angular momentum mode~$n$.

Since the wave packet is localized near infinity, it is preferable to work with
the fundamental solutions~$\grave{\phi}$ having the asymptotics~(\ref{abc2}).
They are related to the functions~$\phi_{1\!/\!2}$ (\ref{phi12}) by
\beq \label{3.3}
\left(\!\begin{array}{cc} {\grave{\phi}}\\
{\overline{\grave{\phi}}}\end{array}\!\right)={\bf{A}}\left(\!\begin{array}{cc} {\phi}_{1}\\
{\phi}_{2}
\end{array}\!\right)
\eeq
where
\[ {\bf{A}} \;:=\;  \left( \!\begin{array}{cc} \alpha + \beta &
i(\alpha - \beta )\\
\overline{\alpha}+\overline{\beta}&
-i(\overline{\alpha}-\overline{\beta})
\end{array}\! \right) . \]
In order to compute the inverse of~${\bf{A}}$, we first note that, from the asymptotics (\ref{abc1})
and (\ref{abc2}) of the fundamental solutions of the radial equation at infinity and at the
event horizon, we obtain the following Wronskians,
\[ w(\,{\grave \phi},{\overline{\grave{\phi}}}\,)\;=\;-2i\omega\;,\quad
w(\,{\acute \phi},{\overline{\acute{\phi}}}\,)\;=\;2i\Omega\;. \]
On the other hand, from (\ref{trans}), we know that
\[ w(\,{\grave
\phi},{\overline{\grave{\phi}}}\,)\;=\;(\,|\alpha|^{2}-|\beta|^{2}\,)w(\,{\acute
\phi},{\overline{\acute{\phi}}}\,)\;. \]
The last two identities imply that
\begin{equation}\label{negdet}
|\alpha|^{2}-|\beta|^{2}\;=\;-\frac{\omega}{\Omega}\;.
\end{equation}
Using this relation, we obtain
\[ {\bf{B}} \;:=\; {\bf{A}}^{-1} \;=\; \frac{\Omega}{2 \omega}
\left( \!\begin{array}{cc}
-\overline{\alpha}+\overline{\beta} & -\alpha+\beta \\
i (\overline{\alpha}+\overline{\beta})& -i (\alpha + \beta)
\end{array}\! \right) \:. \]

Using again the notation~(\ref{vn}), we have
\beq \label{gpsi}
{{\bf{\Psi}}}^{\omega n}_0 \;=\;
\overline{\bf{B}} \grave{\bf{\Psi}}^{\omega n}_0 \:,
\eeq
so we can write the total energy as
\[ \Etot \;=\; \frac{1}{2 \pi}\sum_{n \in \sN}\int_{-\infty}^{\infty}\frac{d\omega}{\omega \Omega}\:
\langle \grave{\bf{\Psi}}^{\omega n}_0,
{\bf{B}}^t {\bf{T}} \overline{\bf{B}} \, \grave{\bf{\Psi}}^{\omega n}_0
\rangle_{\C^2} \:. \]
The quadratic form can be written in components as
\begin{eqnarray}\langle \grave{\bf{\Psi}}^{\omega n}_0,
{\bf{B}}^t {\bf{T}} \overline{\bf{B}} \, \grave{\bf{\Psi}}^{\omega n}_0
\rangle_{\C^2} &=&
(\overline{\grave{\bf{\Psi}}^{\omega n}_0})_1
({\bf{B}}^t {\bf{T}} \overline{\bf{B}})_{11} (\grave{\bf{\Psi}}^{\omega n}_0)_1
\:+\: (\overline{\grave{\bf{\Psi}}^{\omega n}_0})_2
({\bf{B}}^t {\bf{T}} \overline{\bf{B}})_{22} (\grave{\bf{\Psi}}^{\omega n}_0)_2
\label{diagonal} \\
&&+ (\overline{\grave{\bf{\Psi}}^{\omega n}_0})_1
({\bf{B}}^t {\bf{T}} \overline{\bf{B}})_{12} (\grave{\bf{\Psi}}^{\omega n}_0)_2
+ (\overline{\grave{\bf{\Psi}}^{\omega n}_0})_2
({\bf{B}}^t {\bf{T}} \overline{\bf{B}})_{21} (\grave{\bf{\Psi}}^{\omega n}_0)_1
 \:, \quad \label{cross}
\end{eqnarray}
and the matrix~${\bf{B}}^t {\bf{T}} \overline{\bf{B}}$ is given by
\[ {\bf{B}}^t {\bf{T}} \overline{\bf{B}} \;=\; \frac{\Omega}{2 \omega}
\left( \!\begin{array}{cc}
1 & -\overline{\alpha}/\beta \\
-\alpha/\overline{\beta} & 1 \end{array}\! \right) . \]

We consider the contribution by the diagonal terms~(\ref{diagonal})
and the off-diagonal terms~(\ref{cross}) separately.
Re-expressing the diagonal elements in terms of~$\grave{\bf{\Psi}}^{\omega n}_0$ we obtain
\[ (\ref{diagonal}) \;=\;
\frac{\Omega}{2 \omega} \, \langle \grave{\bf{\Psi}}^{\omega n}_0, \grave{\bf{\Psi}}^{\omega n}_0
\rangle_{\C^2} \:, \]
and the corresponding contribution to~$\Etot$ is given by~(\ref{Etotprp}).

It remains to show that the contribution of the off-diagonal terms~(\ref{cross})
to~$\Etot$ tends to zero as~$L \rightarrow \infty$. Intuitively, this can be understood
from the fact that the off-diagonal terms involve oscillatory terms~$e^{\pm i \omega L^2}$,
whose integral becomes small for large~$L$. In order to make the argument precise,
we use an argument based on the Riemann-Lebesgue Lemma.
We only consider the $(1,2)$ element; the argument
for the~$(2,1)$ element is similar. We write
\[ \frac{1}{\omega \Omega} \:(\overline{\grave{\bf{\Psi}}^{\omega n}_0})_1
({\bf{B}}^t {\bf{T}} \overline{\bf{B}})_{12} (\grave{\bf{\Psi}}^{\omega n}_0)_2
\;=\; e^{-2 i \omega L^2} \, g_L(\omega) \]
with
\beq \label{gn}
g_L(\omega) \;=\; -\frac{1}{2} \:\frac{\overline{\alpha}}{\beta}
\left( \frac{1}{\omega}\:
\bra \Psi_0, e^{i \omega L^2} \grave{\Psi}^{\omega n} \ket \right)
\left( \frac{1}{\omega}\:
\bra \overline{e^{i \omega L^2} \grave{\Psi}^{\omega n}}, \Psi_0 \ket \right) \:.
\eeq
For clarity we point out that the $L$-dependence of $g_L$ comes from~$\Psi_0$
(see~(\ref{wavepack})), as well as from the phase factor~$e^{i \omega L^2}$.
The corresponding contribution to~$\Etot$ is obtained by integrating over~$\omega$,
and thus the remaining task is to show that
\[ \lim_{L \rightarrow \infty} \int_{-\infty}^\infty e^{-2 i \omega L^2} \, g_L(\omega)
\: d\omega \;=\; 0\:. \]
This is a direct consequence of the following two lemmas.
\QED

\begin{Lemma} Let~$f_L \in L^1(\R)$, $L>0$, be a family of functions with
which converges in $L^1(\R)$ as~$L \rightarrow \infty$. Then
\[ \lim_{L \rightarrow \infty} \int_{-\infty}^\infty e^{-2 i \omega L^2} \, f_L(\omega)
\: d\omega \;=\; 0\:. \]
\end{Lemma}
{\Proof} Let~$f=\lim_{L \rightarrow \infty} f_L$ in~$L^1$. Using the inequality
\[ \left| \int_{-\infty}^\infty e^{-2 i \omega L^2} \, f_L(\omega)\: d\omega \right|
\;\leq\; \left| \int_{-\infty}^\infty e^{-2 i \omega L^2} \, f(\omega)\: d\omega \right|
+ \int_{-\infty}^\infty |f-f_L|(\omega)\: d\omega \:, \]
the first term tends to zero by the Riemann-Lebesgue lemma, while the second
term tends to zero by hypothesis.
\QED

\begin{Lemma} For any~$n, \tilde{n}$ and~$\tilde{\omega}$, the functions~$g_L$ 
converge in $L^1(\R)$ as~$L \rightarrow \infty$.
\end{Lemma}
{\Proof} In view of Lebesgue's dominated convergence theorem, it suffices to show that
\begin{description}
\item[(a)] The pointwise limit $\lim_{L \rightarrow \infty}g_L(\omega)$ exists for
all~$\omega \neq 0$, and
\item[(b)] There is a constant~$C$ such that for all~$\omega$ and~$L$,
\[ \left| g_L(\omega) \right| \;\leq\; \frac{C}{\omega^2+1} \:. \]
\end{description}
To show~(a), note that in the limit~$L \rightarrow \infty$, the support of~$\Psi_0$
moves to infinity. Thus using the plane wave asymptotics of the Jost solutions~$\psi^{\omega n}$
together with the error estimates in the proof of Lemma~3.3 in~\cite{FKSY2},
a straightforward calculation shows that both brackets in~(\ref{gn}) converge
for any fixed~$\omega \neq 0$. This proves~(a).

To prove~(b), we first analyze the behavior near $\omega=0$. First note that the
factors~$\omega^{-1}$ in the brackets in~(\ref{gn}) are compensated because
the energy scalar product involves a factor of~$\omega$ (see~\cite[Eq.~(2.14)]{FKSY1}).
Thus the estimates for the Jost functions in Lemma~3.6 in~\cite{FKSY2} yield
that the two brackets in~(\ref{gn}) are bounded for small~$|\omega|$, uniformly in~$L$.
Next, the convexity argument in~\cite[Section~5]{FKSY2} yields that the
factor~$\overline{\alpha}/\beta$ is bounded near~$\omega=0$.

On any compact set which does not contain~$\omega=0$, we know from~\cite[Section~7]{FKSY2}
that the factor~$\overline{\alpha}/\beta$ is a continuous function. Since the Jost solutions
form holomorphic families and the estimates of~\cite[Lemma~3.3]{FKSY2} hold, it follows that
the integrand in~(\ref{gn}) is continuous and thus bounded on any compact set, uniformly in~$L$.

To control the behavior for large~$|\omega|$, we first see from~(\ref{negdet}) that~$|\overline{\alpha}/\beta|<1$. Thus to finish the proof of~(b) it suffices to show that the energy scalar
products in~(\ref{gn}) are bounded, uniformly in~$L$.
Indeed, we shall show that they have rapid decay in~$\omega$, uniformly in~$L$, i.e.\
for any~$p \in \N$ there is a constant~$c_p$ such that
\beq \label{rapidestimate}
\left| \bra \Psi_0,  \grave{\Psi}^{\omega n} \ket \right|
+ \left| \bra \overline{ \grave{\Psi}^{\omega n}}, \Psi_0 \ket \right|
\;\leq\; \frac{c_p}{|\omega|^p} \spc \forall\, L,\; \forall\, \omega {\mbox{ with }} |\omega|>1\:.
\eeq
For this we modify the argument in the proof of~\cite[Lemma~3.1~(iii)]{FS}.
We write the radial ODE~(\ref{5ode}) as
\[ \omega^2\, \grave{\phi} \;=\; (-\partial_u^2 + (V+\omega^2))\, \grave{\phi} \:. \]
Iterating this relation gives
\[ \omega^{2l}\, \grave{\phi} \;=\; (-\partial_u^2 + (V+\omega^2))^l\, \grave{\phi} \:. \]
We thus obtain for suitable functions~$F$ and~$G$,
which are independent of~$\omega$, the formula
\[ \omega^{2l}\, \grave{\Psi}^{\omega n} \;=\; (-\partial_u^2 + \omega F + G)^l\,
\grave{\Psi}^{\omega n}\:, \]
where~$\grave{\Psi}^{\omega n} = (\grave{\Phi}^{\omega n}, i \partial_t
\grave{\Phi}^{\omega n})$ and~$\grave{\Phi}^{\omega n}$ is defined by~(\ref{separansatz},
\ref{rescal}). Hence
\begin{eqnarray}
\bra \Psi_0, \grave{\Psi}^{\omega n} \ket &=& \frac{1}{\omega^{2l}}
\: \bra \Psi_0, (-\partial_u^2 + \omega F + G)^{l}\, \grave{\Psi}^{\omega n} \ket \nonumber \\
&=& \frac{1}{\omega^{2l}} \: \bra ((-\partial_u^2 + \omega F + G)^*)^{l}\, \Psi_0,
\grave{\Psi}^{\omega n} \ket\:, \label{nonum}
\end{eqnarray}
where star denotes the formal adjoint obtained by partial integration.
Writing the function~$((-\partial_u^2 + \omega F + G)^*)^l \Psi_0$ as a polynomial in~$\omega$,
each coefficient is again a smooth function in~$u$, which is bounded by~${\mbox{const}}/\sqrt{L}$
and is supported in the interval~$u \in [L^2-L, L^2+L]$.
Hence again using the Jost estimates in~\cite[Section~3.1]{FKSY2}, we conclude
that
\[ \left| \bra ((-\partial_u^2 + \omega F + G)^*)^{l} \Psi_0,
\grave{\Psi}^{\omega n} \ket \right| \;\leq\; C\, |\omega|^{l+1}\:, \]
where the constant~$C$ depends only on~$\Psi_0$ and~$l$, but is independent of~$L$.
Since~$l$ is arbitrary, we obtain the desired estimate for the first summand in~(\ref{rapidestimate}). The proof for the second argument is similar.
\QED
For later use, we restate the inequality~(\ref{rapidestimate}) in terms of
$\grave{{\bf{\Psi}}}^{\omega n}_0$ as defined by~(\ref{psigravcomp}).
\begin{Remark} \label{remark34}
For any~$n, \tilde{n} \in \N$ and~$\tilde{\omega} \in (\omega_0,0)$, 
the function~$\grave{{\bf{\Psi}}}^{\omega n}_0$
has rapid decay in~$\omega$, uniformly in~$L$, i.e.\
for any~$p \in \N$ there is a constant~$c_p$ such that for all~$\omega$ and all~$L$,
\[ \left| \grave{{\bf{\Psi}}}^{\omega n}_0 \right|
\;\leq\; \frac{c_p}{(|\omega|+1)^p} \:. \]
\end{Remark}

The result of Proposition~\ref{prpEtot} can be understood more
directly from the following intuitive considerations. For large~$L$ the wave packet
is localized near infinity, and thus its energy is well-approximated
by the energy in Minkowski space. Considering only one angular mode
and integrating out the angular variables, we thus obtain
\[ \Etot \;=\; \int_{-\infty}^\infty \left(
|\partial_t \phi_0|^2 + |\partial_r \phi_0|^2 \right) du \:+\: {\mathcal{O}}(L^{-1})\:, \]
(where the angular derivatives are contained in the error term).
Writing the solution of the one-dimensional wave equation as a Fourier integral
involving left- and right-going waves,
\[ \phi(t,u) \;=\; \frac{1}{2 \pi} \int_{-\infty}^\infty e^{-i \omega t}
\left( \hat{\phi}_L(\omega)\, e^{-i \omega u} + \hat{\phi}_R(\omega)\, e^{i \omega u} \right) d\omega\:, \]
we can rewrite~$\Etot$ as
\beq \label{Et2}
\Etot \;=\; \frac{1}{\pi} \int_{-\infty}^\infty
\omega^2 \left( |\hat{\phi}_L(\omega)|^2 + |\hat{\phi}_R(\omega)|^2 \right)
d\omega \:+\: {\mathcal{O}}(L^{-1})\:.
\eeq
Next, using the asymptotics~(\ref{abc2}) of the fundamental solutions~$\grave{\phi}$
together with the asymptotic form of the energy density in~(\ref{energysp}), we find that
\[ \hat{\phi}_L(\omega) \;=\; \frac{1}{2 \omega^2}\: \bra \grave{\Psi}^{\omega n}, \Psi_0 \ket
\:+\: {\mathcal{O}}(L^{-1})\:,\quad
\hat{\phi}_R(\omega) \;=\; \frac{1}{2 \omega^2}\: \bra \overline{\grave{\Psi}^{\omega n}}, \Psi_0 \ket
\:+\: {\mathcal{O}}(L^{-1})\:. \]
Using this formula in~(\ref{Et2}), we again get the expression in Proposition~\ref{prpEtot}.

Using the same method as in Proposition~\ref{prpEtot}, we
next compute the outgoing energy~$\Eout$ for our wave packet.
\begin{Prp} \label{prpEout}
For the initial data given by the wave packet~(\ref{wavepack}),
\beq \label{Eoutprp} \lim_{L \rightarrow \infty} \Eout(\Psi_0)
\;=\; \frac{1}{4 \pi} \sum_{n \in \sN} \lim_{L \rightarrow \infty}
\int_{-\infty}^{\infty}\frac{d\omega}{\omega^2}\: \langle
\grave{\bf{\Psi}}^{\omega n}_0, {\bf{R}}\,
\grave{\bf{\Psi}}^{\omega n}_0
\rangle_{\C^2}\:, \eeq
where~${\bf{R}}$ is the matrix
\beq \label{Rdef}
{\bf{R}} \;=\; \left(\!\begin{array}{cc} \displaystyle
\frac{|\alpha|^2}{|\beta|^2} & 0 \\[.8em] 0 & 1
\end{array}\! \right) .
\eeq
The above series converges absolutely.
\end{Prp}
{\Proof} Applying~\cite[Theorem~8.1]{FS}, we see that
the energy near infinity of the large angular momentum modes is small uniformly in~$L$.
This allows us to interchange the limit~$L \rightarrow \infty$ with the
sum over the angular momentum modes. Hence in what follows
we may restrict attention to a fixed angular momentum mode~$n$.

Using~(\ref{gpsi}), we can write the result of
Proposition~\ref{Eoutexpr} as
\[ \Eout \;=\; \frac{1}{2 \pi}\sum_{n \in \sN}\int_{-\infty}^{\infty}\frac{d\omega}{\omega \Omega}\:
\langle \grave{\bf{\Psi}}^{\omega n}_0,
{\bf{B}}^t {\bf{T}} {\bf{Q}} {\bf{T}} \overline{\bf{B}} \,
\grave{\bf{\Psi}}^{\omega n}_0 \rangle_{\C^2} \:. \]
The matrix~${\bf{B}}^t {\bf{T}} {\bf{Q}} {\bf{T}} \overline{\bf{B}}$
can be computed to be
\[ {\bf{B}}^t {\bf{T}} {\bf{Q}} {\bf{T}} \overline{\bf{B}}
\;=\; \frac{\Omega}{2 \omega}
\left(\!\begin{array}{cc} \displaystyle
|\alpha|^2/|\beta|^2 & -\overline{\alpha}/\beta \\
\alpha/\overline{\beta} & 1
\end{array}\! \right) . \]
The diagonal terms give rise to~(\ref{Eoutprp}). The off-diagonal
terms, on the other hand, are exactly the same as
those considered in Proposition~\ref{prpEtot}. Thus we already know
that they vanish in the limit~$L \rightarrow \infty$.
\QED

\section{Computation of Energy Gain}\label{energain}
\setcounter{equation}{0}
Combining Propositions~\ref{prpEtot} and~\ref{prpEout}, we obtain
\beq
\label{preRayleigh} \lim_{L \rightarrow \infty}
\frac{\Eout}{\Etot} \;=\;
\frac{ \displaystyle \sum_{n \in \N} \lim_{L \rightarrow \infty} 
\int_{-\infty}^\infty \lbra \grave {\bf{\Psi}}^{\omega n}_0,
{\bf{R}}(\omega, n)\, \grave {\bf{\Psi}}^{\omega
n}_0 \lket_{\C^2}\,\frac{ d\omega}{\omega^{2}}} {\displaystyle
\sum_{n \in \N} \lim_{L \rightarrow \infty} \int_{-\infty}^\infty \lbra \grave
{\bf{\Psi}}^{\omega n}_0, \grave {\bf{\Psi}}^{\omega n}_0
\lket_{\C^2}\, \frac{d\omega}{\omega^{2}}} \:.
\eeq
To discuss this formula, let us see how to recover
Starobinsky's result~(\ref{frakRdef}). To this end, we
replace the superposition in the numerator and denominator
by a single wave mode with quantum numbers $\tilde{\omega} \neq 0$
and~$\tilde{n}$. Then~(\ref{preRayleigh}) simplifies to
\[ \frac{\Eout}{\Etot} \;=\; \frac{\lbra \grave {\bf{\Psi}}^{\tilde{\omega}
\tilde{n}}_0,
{\bf{R}}(\tilde{\omega}, \tilde{n})\, \grave {\bf{\Psi}}^{\tilde{\omega}
\tilde{n}}_0 \lket_{\C^2}}
{\lbra \grave {\bf{\Psi}}^{\tilde{\omega} \tilde{n}}_0, \grave {\bf{\Psi}}^{\tilde{\omega} \tilde{n}}_0
\lket_{\C^2}} \:. \]
If we consider an outgoing wave at infinity, the
vector~$\grave {\bf{\Psi}}^{\tilde{\omega} \tilde{n}}_0$
vanishes in the first component. According to~(\ref{Rdef}),
the above quotient gives one. This result is immediately clear,
because if we take an outgoing wave near infinity, it will not
interact with the black hole and simply escape to infinity.
It is more interesting to consider an incoming wave at infinity.
In this case, the second component of the vector~$\grave
{\bf{\Psi}}^{\tilde{\omega} \tilde{n}}_0$ vanishes, and thus
\[ \frac{\Eout}{\Etot} \;=\; 
\left| \frac{\alpha(\tilde{\omega})}{\beta(\tilde{\omega})} \right|^2
\:. \]
Expressing~$A$ and~$B$ in terms of our transmission
coefficients~$\alpha$ and~$\beta$ (by a straightforward
calculation using~(\ref{decompphi}, \ref{trans}) together
with~(\ref{negdet})), this is in complete agreement with
Starobinsky's result~\cite{S}.

The remaining task is to prove that in the limit~$L \rightarrow \infty$,
our wave packets become ``more and more localized''
near~$\omega=\tilde{\omega}$ and~$n=\tilde{n}$. \\[.5em]
{\em{Proof of Theorem~\ref{maintheo}. }} We consider the wave
packet initial data~(\ref{wavepack}) with~$\tilde{\omega}$
and~$\tilde{n}$ equal to $\omega$ and $n$ in the statement of the
theorem. We choose~$\cout=0$ and~$\cin=1$.
From the asymptotic form of the energy density near infinity,
it is obvious that the total energy of the wave packet~(\ref{wavepack})
has a non-zero limit as~$L \rightarrow \infty$.
Hence the denominator in~(\ref{preRayleigh}) is non-zero.
Moreover, since both series in~(\ref{preRayleigh}) converge absolutely,
they may be approximated by finite sums.
Using the identity
\[ \frac{\Eout}{\bra \Psi_0, \Psi_0 \ket}
- {\mathfrak{R}} \;=\; \frac{1}{\bra \Psi_0, \Psi_0 \ket}
\left( \Eout - \left| \frac{\alpha(\tilde{\omega})}{\beta(\tilde{\omega})} \right|^2
\bra \Psi_0, \Psi_0 \ket \right) \]
and considering a finite number of modes,
we see that our task is to show that
\beq \label{finmod} \lim_{L \rightarrow \infty} \sum_{n \leq n_0}
\int_{-\infty}^\infty \lbra \grave {\bf{\Psi}}^{\omega n}_0,
\left( {\bf{R}}(\omega,n) - \left| \frac{\alpha(\tilde{\omega})}
{\beta(\tilde{\omega})} \right|^2 \right) \grave
{\bf{\Psi}}^{\omega n}_0 \lket_{\C^2}\, \frac{d\omega}{\omega^{2}}
\;=\; 0 \:. \eeq
It clearly suffices to prove this for one mode.

In the case~$n=\tilde{n}$, for given~$\varepsilon>0$ we choose an
open neighborhood~$I_\varepsilon$ of~$\tilde{\omega}$ such that
\[ \frac{1}{\omega^{2}} \left|  \left| \frac{\alpha(\omega)} {\beta(\omega)} \right|^2 - \left| \frac{\alpha(\tilde{\omega})} {\beta(\tilde{\omega})} \right|^2 \right| 
\;\leq\; \varepsilon \spc {\mbox{for all }} \omega \in I_\varepsilon\:. \]
Using the asymptotic form of the energy density and of the
fundamental solutions, a short computation shows that
\begin{eqnarray*}
\lim_{L \rightarrow \infty} (\grave {\bf{\Psi}}^{\omega \tilde{n}}_0)_2
&=& 0 \quad {\mbox{pointwise in $\R$}} \\
\lim_{L \rightarrow \infty} \grave {\bf{\Psi}}^{\omega \tilde{n}}_0 &=& 0
\quad {\mbox{pointwise in $\R \setminus I_\varepsilon$}}
\end{eqnarray*}
(where~$(\grave {\bf{\Psi}}^{\omega \tilde{n}}_0)_2$ denotes the second component
of~$\grave {\bf{\Psi}}^{\omega \tilde{n}}_0$), and that
the functions~$(\grave {\bf{\Psi}}^{\omega \tilde{n}}_0)_2$ and
$\grave {\bf{\Psi}}^{\omega \tilde{n}}_0$ are dominated in~$\R$
resp.\ in~$\R \setminus I_\varepsilon$ by Schwartz functions, uniformly in~$L$
(see Remark~\ref{remark34}).
Hence applying Lebesgue's dominated convergence theorem, we obtain
\begin{eqnarray*}
\lim_{L \rightarrow \infty} \int_{I_\varepsilon} \lbra \grave {\bf{\Psi}}^{\omega \tilde{n}}_0,
\left({\bf{R}}(\omega,\tilde{n}) - \left| \frac{\alpha(\tilde{\omega})}{\beta(\tilde{\omega})} \right|^2 \right)
\, \grave {\bf{\Psi}}^{\omega \tilde{n}}_0 \lket_{\C^2}\,
\frac{d\omega}{\omega^{2}} &\leq& \varepsilon\: \Etot \\
\lim_{L \rightarrow \infty} \int_{\R \setminus I_\varepsilon} \lbra \grave {\bf{\Psi}}^{\omega \tilde{n}}_0,
\grave {\bf{\Psi}}^{\omega \tilde{n}}_0 \lket_{\C^2}\,
\frac{d\omega}{\omega^{2}} &=& 0\:.
\end{eqnarray*}
Since~$\varepsilon$ is arbitrary, this proves~(\ref{finmod})
for the summand~$n=\tilde{n}$.

If~$n \neq \tilde{n}$, the orthogonality of the spheroidal wave
functions for $\omega=\tilde{\omega}$ together with the continuity
in~$\omega$ allows us to choose a neighborhood~$I_\varepsilon$
of~$\tilde{\omega}$ such that
\[ \bra \Theta^{n, \omega}, \Theta^{\tilde{n}, \tilde{\omega}} \ket_{S^2} \;\leq\; \varepsilon
\spc {\mbox{for all }} \omega \in I_\varepsilon\:. \]
Now we can use the same argument as in the case~$n =\tilde{n}$ to conclude the
proof of~(\ref{finmod}).
\QED

\section{Absorbtion of Angular Momentum by the Black Hole}
\setcounter{equation}{0} We first derive the expression for the
angular momentum of the scalar wave. We recall from~\cite{FKSY1}
that the Lagrangian is given by
\begin{eqnarray*}
{\cal{L}}&=&-\Delta|\partial_{r}\Phi|^{2}+\frac{1}{\Delta} \left|((r^{2}+a^{2})\partial_{t}
+a\partial_{\varphi})\Phi \right|^{2} \nonumber \\
&& -\sin^{2}{\vartheta} \left|\partial_{\cos \vartheta}\varphi \right|^{2} -\frac{1}{\sin^{2}\vartheta} \left| (a\sin^{2}\vartheta\partial_{t}+\partial_{\varphi})\Phi \right|^{2} .
\end{eqnarray*}
This Lagrangian is axisymmetric. Applying Noether's theorem gives rise to the
following conserved quantity,
\[ A[\Phi] \;=\; \int_{r_1}^{\infty} dr \int_{-1}^{1}d(\cos \vartheta)
\int_{0}^{2\pi}\frac{d\varphi}{2 \pi} \:{\cal{A}}\,, \]
where ${\cal{A}}$ is given by
\begin{eqnarray*}
{\cal{A}} &=& {\mbox{Re}} \left( \frac{\partial {\cal L}}{\partial \Phi_{t}}\,\Phi_\varphi \right) \\
&=& {\mbox{Re}} \left\{
\frac{(r^{2}+a^{2})^{2}}{\Delta}\:
\overline{\partial_\varphi \Phi} \left(\partial_{t}\Phi+\frac{a\:\partial_\varphi \Phi}{r^{2}+a^{2}} \right)
-a^{2}\sin^{2}\vartheta \:\overline{\partial_\varphi \Phi}
\left(\partial_{t}\Phi+\frac{\partial_{\varphi}\Phi}{a
\sin^{2}\vartheta}\right) \right\} .
\end{eqnarray*}
The quantity $A$ can be interpreted as the angular momentum of the wave~$\Phi$,
and~${\mathcal{A}}$ as the angular momentum density.

Similar to~(\ref{Eindef}), the angular momentum absorbed by the black hole
is defined by
\beq \label{Mbhdef}
\Mbh \;=\; \lim_{t \to \infty}
\int_{r_1}^{2 r_1} dr \int_{-1}^{1}d(\cos \vartheta)
\int_{0}^{2\pi}\frac{d\varphi}{2 \pi} \:{\cal{A}}[\Phi] \:,
\eeq
provided that the limit exists.
In the next proposition we compute~$\Mbh$, again for a fixed $k$-mode.
\begin{Prp}\label{Minexpr}
Choose a fixed $k\in \Z$ such that $\omega_{0}<0$. Then
the limit in~(\ref{Mbhdef}) exists and
\[ \Mbh \;=\; \frac{1}{2 \pi}\sum_{n \in \sN}\int_{-\infty}^{\infty}\frac{d\omega}{\omega \Omega}\:
\frac{k}{\Omega}
\lbra {\bf{\Psi}}_{0}^{\omega n},{\bf T}{\bf P}{\bf T} \,{\bf{\Psi}}_{0}^{ \omega
n}\lket_{\C^{2}}\:, \]
where the sum converges absolutely.
\end{Prp}
{\Proof} As in the proof of Proposition~\ref{prpEout} one sees that we may interchange
the limit~$t \rightarrow \infty$ with the sum over the angular momentum modes,
and that the resulting series converges absolutely. Thus it remains to consider
a fixed angular momentum mode~$n$.
Substituting the integral representation~(\ref{intrep}) into the definition
of~$\Mbh$, we obtain
\[ \Mbh \;=\; \lim_{t\to
\infty}\frac{1}{(2\pi)^{2}}
\int_{r_1}^{2 r_1} dr \int_{-\infty}^{\infty}\frac{d\omega'}{\omega' \Omega'}
\int_{-\infty}^{\infty}\frac{d\omega}{\omega \Omega} \:e^{-i (\omega-\omega') t}
\:{\mathcal{A}}^{\omega', \omega}(r) \;, \]
where
\[ {\mathcal{A}}^{\omega', \omega}(r) \;=\; \sum_{n,n' \in \N}
\sum_{a,b,c,d=1}^2 t^{\omega' n'}_{ab} t^{\omega n}_{cd}\:
\bra \Psi_0, \Psi^{\omega' n'}_a \ket
\: \bra \Psi^{\omega n}_d, \Psi_0 \ket
\int_{S^2} {\mathcal{A}}(\Psi^{\omega' n'}_b, \Psi^{\omega n}_c)\:, \]
and~${\mathcal{A}}(\Psi^{\omega' n'}_b, \Psi^{\omega n}_{c})$ denotes
the bilinear form corresponding to the angular momentum density
(similar to the bilinear form~${\mathcal{E}}(\Psi^{\omega' n'}_b, \Psi^{\omega n}_{c})$
appearing in the proof of Proposition~\ref{Einexpr}).
Near the event horizon, we can expand~${\mathcal{A}}(\Psi^{\omega' n'}_b,
\Psi^{\omega n}_{c})$ to obtain
\begin{eqnarray*}
\frac{\Delta}{r^2+a^2}\: {\mathcal{A}}(\Psi^{\omega' n'}_b, \Psi^{\omega n}_{c})
&=& \frac{1}{2}\:(r_1^2+a^2)\:
k \,(\Omega+\Omega')\:\overline{\Phi^{\omega' n'}_b}\, \Phi^{\omega n}_c
\:+\: {\mathcal{O}}(e^{\gamma u}) \\
&=& \frac{1}{2}\:\Theta_{n'\omega'} \Theta_{n \omega}\:
k \,(\Omega+\Omega')\: \overline{\phi^{\omega' n'}_b}\, \phi^{\omega n}_c \:+\: {\mathcal{O}}(e^{\gamma u}) .
\end{eqnarray*}
Now we can proceed exactly as in the proof of Proposition~\ref{Einexpr}.
\QED

\noindent
{\em{Proof of Theorem~\ref{thm2}. }} Without loss of generality we again
restrict attention to the case~$k>0$, so that~$\omega_0<0$. We set
\[ \rho(\omega, n) \;=\; \frac{1}{2 \pi}\:
\lbra {\bf{\Psi}}_{0}^{\omega n},{\bf T}{\bf P}{\bf T} \,{\bf{\Psi}}_{0}^{ \omega
n}\lket_{\C^{2}} \:. \]
The eigenvalues of the matrix~${\bf T}{\bf P}{\bf T}$ are computed to be zero
and~$1 + |\alpha|^2/|\beta|^2$. Thus~$\rho$ is non-negative. It follows that
\begin{eqnarray*}
\Mbh &=& \sum_{n \in \N} \int_{-\infty}^\infty \frac{k}{\omega \Omega^2}\:\rho\:d\omega
\;\leq\; \sum_{n \in \N} \int_{\omega_0}^\infty \frac{k}{\omega \Omega^2}\:\rho\:d\omega \\
&\leq& \frac{k}{|\omega_0|} \sum_{n \in \N} \int_{\omega_0}^\infty \frac{1}{\omega \Omega}\:\rho\:d\omega
\;\leq\; \frac{k}{|\omega_0|} \sum_{n \in \N} \int_{-\infty}^\infty \frac{1}{\omega \Omega}\:\rho\:d\omega
\;=\; \frac{k}{|\omega_0|}\: \Ebh\:,
\end{eqnarray*}
and using~(\ref{Odef}) completes the proof.
\QED

\noindent
{\em{Acknowledgments:}}
We would like to thank the Alexander-von-Humboldt Foundation and the Vielberth
Foundation, Regensburg, for generous support. We also thank the referee
his careful reading and helpful suggestions.

\begin{tabular}{lcl}
\\
Felix Finster & $\;\;\;\;$ & Niky Kamran\\
NWF I -- Mathematik && Department of Math.\ and Statistics \\
Universit{{\"a}}t Regensburg && McGill University \\
93040 Regensburg, Germany && Montr{\'e}al, Qu{\'e}bec \\
{\tt{Felix.Finster@mathematik}} && Canada H3A 2K6  \\
$\;\;\;\;\;\;\;\;\;\;\;\;\;\;$ {\tt{.uni-regensburg.de}}
&& {\tt{nkamran@math.McGill.CA}} \\
\\
Joel Smoller & $\;\;$ & Shing-Tung Yau \\
Mathematics Department && Mathematics Department \\
The University of Michigan && Harvard University \\
Ann Arbor, MI 48109, USA && Cambridge, MA 02138, USA \\
{\tt{smoller@umich.edu}} && {\tt{yau@math.harvard.edu}}
\end{tabular}


\addcontentsline{toc}{section}{References}
\begin{thebibliography}{99}
\bibitem{ALP} N.\ Andersson, P.\ Laguna, P.\ Papadopoulos,
``Dynamics of scalar fields in the background of rotating black
holes II: a note on superradiance,'' gr-qc/9802059, {\em{Phys. Rev.}}~{\bf{D58}}
(1998) 087503.
\bibitem{Ch} S.\ Chandrasekhar, ``The Mathematical Theory of Black
Holes,'' {\em{Oxford University Press}} (1983).
\bibitem{Christo} D.\ Christodoulou, ``Reversible and irreversible
transformations in black hole physics,''{\em Phys. Rev. Lett.} {\bf
25}, 1956-1957.
\bibitem{DR1} N.\ Deruelle, R.\ Ruffini, ``Quantum and classical relativistic energy
states in stationary geometries,'' {\em{Phys.\ Lett.}}\ {\bf{52B}} (1974) 437
\bibitem{DR2} N.\ Deruelle, R.\ Ruffini, ``Klein Paradox in a Kerr Geometry,''
{\em{Phys.\ Lett.}}\ {\bf{57B}} (1975) 248
\bibitem{FKSY03} F.\ Finster, N.\ Kamran, J.\ Smoller and S.-T.\ Yau,
``The long-time dynamics of Dirac particles in the Kerr-Newman black
hole geometry,'' gr-qc/0005088,
{\em{Adv.\ Theor.\ Math.\ Phys.}}\ {\bf 7} (2003) 25--52.
\bibitem{FKSY02} F.\ Finster, N.\ Kamran , J.\ Smoller and S.-T.\ Yau,
``Decay rates and probability estimates for masssive Dirac particles
in the Kerr-Newman black hole geometry,'' gr-qc/0107094, {\em Commun.\ Math.\ Phys.}\
{\bf 230} (2002) 201--244.
\bibitem{FKSY1} F.\ Finster, N.\ Kamran, J.\ Smoller and S.-T.\ Yau,
``An integral spectral representation of the propagator for the wave
equation in the Kerr geometry,'' gr-qc/0310024, {\em{Commun.\ Math.\ Phys.}}\ {\bf 260}
(2005) 257--298.
\bibitem{FKSY2} F.\ Finster,  N.\ Kamran, J.\ Smoller and S.-T.\ Yau,
``Decay of solutions of the wave equation in the Kerr
geometry,'' gr-qc/0504047, {\em{Commun.\ Math.\ Phys.}}\ {\bf 264} (2006) 465--503,
erratum {\em{Commun.\ Math.\ Phys.}}\ {\bf 280} (2008) 563--573
\bibitem{FS} F.\ Finster, J.\ Smoller, ``A time independent energy estimate for
outgoing scalar waves in the Kerr geometry,'' arXiv:0707.2290 [math-ph],
{\em{J.\ Hyperbolic Differ.\ Equ.}}\ {\bf{5}} (2008) 221--255
\bibitem{FN} V.P.\ Frolov, I.D.\ Novikov,
``Black Hole Physics. Basic Concepts and New Developments,'' {\em
Kluwer Academic Publishers Group, Dordrecht} (1998).
\bibitem{Penrose} R.\ Penrose, ``Gravitational collapse: The role of
general relativity,'' {\em{Rev.\ del Nuovo Cimento}} {\bf 1} (1969)
252--276.
\bibitem{S} A.A.\ Starobinsky, ``Amplification of waves during reflection
from a black hole,'' {\em{Soviet Physics JETP}} {\bf{37}} (1973)
28--32.
\bibitem{TeP} S.\ Teukolsky, W.H.\ Press, ``Perturbations of a
rotating black hole. III. Interaction of the hole with gravitational
and electromagnetic radiation,'' {\em{Astrophys.\ J.}}\ {\bf 193}
(1974) 443--461.
\bibitem{Wald} R.\ Wald, ``General Relativity,'' {\em{University of Chicago
Press}} (1984).
\bibitem{Z} Ya.B.\ Zel'dovich, ``Amplification of cylindrical electromagnetic
waves from a rotating body,'' {\em{Soviet Physics JETP}} {\bf{35}}
(1972) 1085--1087.
\end{thebibliography}
\end{document}